# Evaluating the Possibility of Integrating Augmented Reality and Internet of Things Technologies to Help Patients with Alzheimer's Disease

Fatemeh Ghorbani, Mohammad Kia, Mehdi Delrobaei
Faculty of Electrical Engineering and the Center for Research and Technology (CREATECH)
K. N. Toosi University of Technology
Tehran, Iran
fatemeghorbani@email.kntu.ac.ir, kamran@email.kntu.ac.ir, delrobaei@kntu.ac.ir

Quazi Rahman
Innovation Centre for Information Engineering
Department of Electrical and Computer Engineering
Western University
London, ON, Canada
qrahman3@uwo.ca

*Abstract*— **People suffering from Alzheimer's disease (AD) and their caregivers seek different approaches to cope with memory loss. Although the AD patients want to live independently, they often need help from the caregivers. In this situation, caregivers may attach notes on every single object or taking out contents of drawer to make them visible before leaving the patient alone at home. This study reports preliminary results on an Ambient Assisted Living (AAL) real-time system, achieved through Internet of Things (IoT) and Augmented Reality (AR) concepts, aimed at helping people suffering from AD. The system has two main sections: the first one is the smartphone or windows application that allows caregivers to monitor patients' status at home and be notified if patient are at risk. The second part allows patient to use smart glasses to recognize QR codes in the environment and receive information related to tags in the form of audio, text or three-dimensional image. This work presents preliminary results and investigates the possibility of implementing such a system.**

*Keywords: Augmented Reality; Internet of Things; Alzheimer's Disease; Ambient Assisted Living; Distributed Mechatronics*

## I. Introduction

Alzheimer's disease (AD) is the main common cause of dementia. Approximated 5.8 million Americans are suffering from AD in 2019. By 2025, the number of people over age 65 with AD is predicted to achieve 7.1 million [1]. Loss of memory is the main difficulties experienced by patients with AD. In fact, they have problems with remembering recent information. Thus, they need to be continually reminded of tasks to be performed that have a considerable effect on their confidence and quality of life. This can disrupt patients and their family's daily life. In this situation, caregivers use their best architectural design to create simple environment. For instance, they use clear plastic boxes to make objects more visible or leave reminder notes (e.g., "Take your medicine at noon" or "Don't leave the home") although these methods have not been entirely effective in most cases [2].

Recently, Augmented Reality (AR) has been used as a cognitive aid tool to help people suffering from AD, specifically in the early stages [3]. In early stages, they may experience memory failure, such as forgetting how to deal with routine tasks or the location of their belongings [4]. The ARCoach system has been designed by researcher at Chung Yuan Christian University, which was based on using personal computer with an external web camera to add information to a real object [5]. The Ambient aNnotation System (ANS), was included two main section, first one allowed the caregiver to create new AR tags by selecting objects or locations then annotating them with information that could help the person with AD. A second one ran on a smartphone was used by person with Alzheimer's disease and alerted of the existence of tags in the environment [6]. Despite their useful features, these systems could not be used in daily life easily, furthermore they did not have any monitoring option, so family consternation still remained.

During the last years, development of Internet of Things (IoT) services have drawn a considerable quantity of concentration in the scientific association. However, few researchers have addressed the development of a model designed for seniors with chronic diseases and particular demands, such as dementia and AD. Ambient assisted living (AAL) is an IoT-based service that supports care of elderly or debilitated patients [7]. AAL not only contributes a safer environment but also provides independency and encourages the user to be more physically and mentally active. According to [8] AAL has achieved the first rank among IoT various applications in health care and has had the potential to be a reliable opportunity in the near future.

In contrast with these works, our aim is to provide a real-time AAL system and improve AD patients' ability to carry out everyday task on their own, in-home by designing the system based on smart glasses, without any interference of patients' privacy. The main aim of this study is introducing preliminary results of possibility for this implementation. Despite this interest, no one to the best of our knowledge has studied on integration of AR and IoT technologies to develop an effective AAL system for interacting with patients and monitoring them via Internet by family. We have also considered many parameters such as availability and reliability of the system to implement it.

## II. INTERNET OF THINGS IMPLEMENTATION

### A. MQTT Middleware

The Organization for the Advancement of Structured information (OASIS) approved Message Queuing Telemetry Transport (MQTT) Version 3.1.1 as an OASIS Standard in 2014, and it has been defined as the reference standard for IoT [9]. MQTT is a publish-subscribe message transfer protocol that includes managing communication unit (broker) and clients. The MQTT broker is a message transfer platform that authorizes the message producer client to publish messages with a message identifier topic. Fig. 1 shows MQTT protocol operation in our study. Each client subscribes to one or more topics, a client which sends messages, called publisher and another one that receives messages is subscriber. Clients must notice hostname/IP and port of the broker in order to publish-subscribe to messages. In this study, we used this standard because the message header requires only 2 bytes, so it is extremely lightweight publish/subscribe messaging protocol for small and constrained devices. We created both of our application for patient and his family based on Android operating system. We have also designed another user interface, windows application, for patient's family to show the notifications related to the events detected by the sensors and to control other actuators states in the home environment.

### B. Sensors and actuators

For monitoring patients and making interaction with them, five types of sensors and actuators were placed in four different locations including bedroom, kitchen, TV room and the main entrance. We also used QR codes for reminding regular events according to sensors and actuators value. The general architecture of the designed AAL system is shown in Fig. 2.

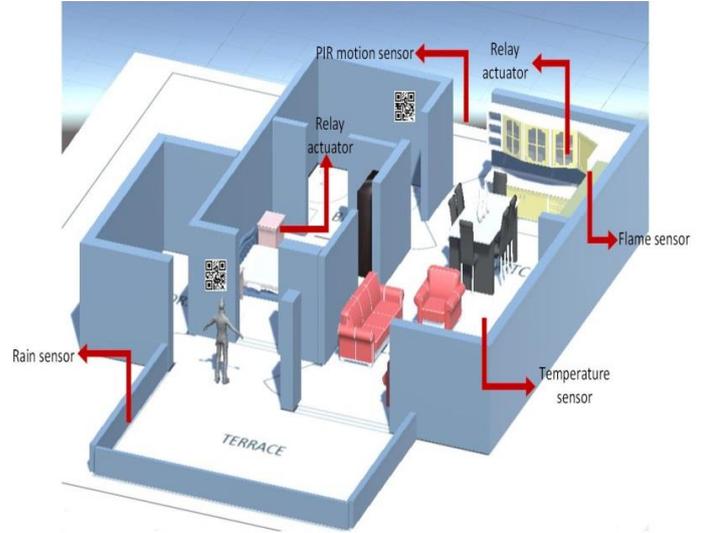

Fig. 2. General architecture of AAL system.

We used rain sensor in the terrace for rain detection, enabling the system to notify the user based the event in the form of AR messages. For fire detection, we considered flam sensor in the kitchen. Moreover, for reminding the weather condition changing, we utilized temperature sensor in the TV room as shown in Fig. 2. We used relay actuators to open or close the drawers or cabinet doors based on certain events, which we describe in Section IV. In addition, PIR motion sensor was used for recognizing user presence in the kitchen.

We attached QR codes on the specific places, where user probably staring at them, for giving some extra information to the user. For example, on the bedroom door, the user could view his family picture and get relevant clues about them in the form of an audio message. These messages could be helpful for memory difficulties.

### C. Platform

The ESP8266-12 [10] boards provide communication abilities to the sensors, through a built-in Wi-Fi interface. Thus, by means of a Wi-Fi home Access Point (AP), sensors are allowed to connect straightly to the broker and publish messages whenever an event is detected via MQTT protocol. Once detected a sensor activation, the node that generally remains in a low power consumption mode starts the Wi-Fi connection phase. When the connection is recognized, the node is able to access the remote MQTT broker, as a result the MQTT connection begins, supposing to the role of MQTT client, and once connected, it publishes a message on the established topic [11]. In our case, sensors were connected to ESP8266 in order to publish messages for example temperature, gas, flame etc. values, and family from a smartphone or a computer could read this value by subscribing to this topic. When the smartphone was offline, all the notifications could be stored by the server and then sent again when the smartphone connected to the Internet. Family could also publish a value such as LED or relay's statues and the ESP8266 needed to subscribe to this value in order to read it. The publish-subscribe messaging pattern requires a message broker. Broker is used to manage and handle all these messages between all subscribed clients. We used several WEMOS D1 mini with a DHT shield to send the temperature, humidity, motions, etc. to a MQTT topic and applied HiveMQ as a MQTT messages broker on ESP8266.

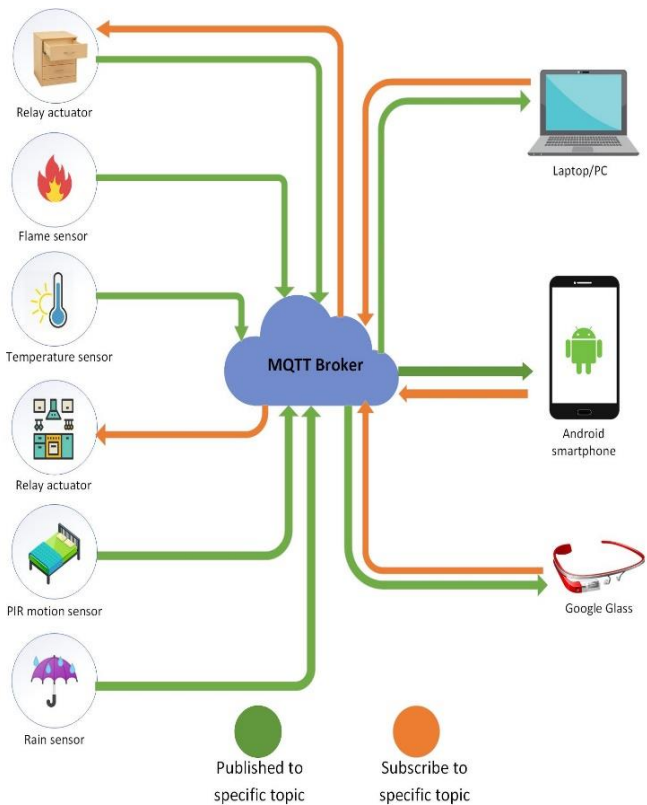

Fig. 1. MQTT protocol operation in AAL system

### III. AUGMENTED REALITY APPLICATION

For designing our AR application, we have used Android ARCore platform, which is Google's open source AR Software Development Kit (SDK) and it is capable of creating three-dimensional contents. In this paper, we developed our application based on android smartphone, and then we investigated the possibility of transferring this app on the Google Glass as the ultimate tool. In this section, we first explain why we have chosen Google Glass as a user interface for AD patient, then we consider the difference between AR and QR tags in AR applications. Finally, we perform our current application on a smartphone and present preliminary results for system implementation.

#### A. Google Glass

Google Glass has clear utility in the clinical setting, i.e. surgery [12], assistive device for people with Parkinson's [13], remote chest X-ray interpretation [14], surgical education [15], vital signs monitoring [16], patient monitoring [17], etc. It also has applications in robotics, i.e. remote control of a mobile robot [18]. We have chosen Google Glass for AR implementation as a future study because of its minimal heads-up display and lightweight, so it could be used by each mild to moderate AD patient simply after short-term training and does not need a user with special skills to navigate it. In compare to other AR glasses, it gives patients access to information in the simplest way that does not distract their daily life and it makes enough essential virtual information while patient is interacting with physical world. Moreover, if patient is a glass-wearer, real glasses can have Google Glass screwed onto them. For example, Microsoft HoloLens is too bulky and distracting for patient, so it is not appropriate for this purpose. We aware that our application could run on Google Glass, because the glass has the same operating system as Android smartphone.

Table 1 classifies the specification of the Google Glass which includes common components found in smart devices, i.e. central processing unit (CPU), camera, global positioning system (GPS), speaker, microphone, display, etc.

#### B. QR Codes

In this paper, we have used six different QR codes, because they can be simply processed using free open source software. In addition, QR codes printed on paper are low cost. However, they can only be recognized one at a time, which was not limiting in this study because we required simultaneous recognition of single symbols. On the other hand, if we want to recognize multiple symbols in future study, we must use AR tags. Similar to QR codes, the AR tags can be printed on paper using open source software [5]. Therefore, it is low cost to make.

TABLE I. GOOGLE GLASS SPECIFICATIONS

| Specifications | |
|---|---|
| Processor | Dual-core 1.2-GHz Texas Instruments OMAP 4430 SoC with Power VR SGX540 GPU |
| Connectivity | Bluetooth 4.0 and Wi-Fi 802.11 b/g |
| Display | 640*360 resolution |
| Storage | 16 GB memory with Google cloud storage |
| Camera | 5.0-megapixel camera, capable of 720p video recording with 30 frames per second rate |
| Battery | 570mAh lithium-polymer batt |
| Charger | Micro USB and charger (outlet or PC charging) |
| Weigh | 50 gr |
| Compatibility | Motion Process Library (MPL) Accelerometer, MPL Gyroscope, MPL Magnetic Field, MPL Orientation, MPL Rotation Vector, MPL Linear Acceleration, MPL Gravity |
| Sensors | LTR-506ALS Light Sensor<br>Rotation Vector Sensor<br>Gravity Sensor<br>Linear Acceleration Sensor<br>Orientation Sensor<br>Corrected Gyroscope Sensor |

In some scenarios, three second timeout occurred in QR code detection algorithm when we wanted to recognize user was staring at QR code.

#### C. Software implementation on the smartphone

For simulation and evaluation of our AAL system, we have used Samsung Galaxy S7 smartphone, which has quad-core Snapdragon 820 processor, 4GB RAM and 12MP rear camera. Fig. 3 shows some pictures based on scenarios we have defined of our distributed mechatronics system. For example, in Fig. 3(a) rain warning was detected by the rain sensor and at the same time user was staring at the main entrance, so after this event they could see an umbrella and hear an audio message which was reminding him some information. In Fig. 3(b), flame sensor in the kitchen near the oven detected the presence of a flame, so relay actuator was activated and user received an image notification. User could also observe the current and previous state image by scrolling. In Fig. 3(c) this state is shown on the LCD screen. The dishes picture for reminding the place of them in the kitchen, shows the current state, which is based on the user's field of view, and the right image is the pervious picture that was sent to the user by family to remind them of the medication time.

### IV. SCENARIOS OF USE

We are considering for possibility of using this system in daily life of AD patient. However, in this paper, we have limited to explain some distributed mechatronics scenarios, which were evaluated in next section. The following sample scenarios describe how the AAL system would be used as a memory aid:

*1) When relay actuator is activated by the smartphone application, the drawer lock is opened and user can see an image of pills on the screen. In addition, user receives an audio message to remind them of the medication time.*

*2) When the QR code attached on the bedroom door is detected by the camera, user can see their family picture and get some personal information about them as an audio message.*

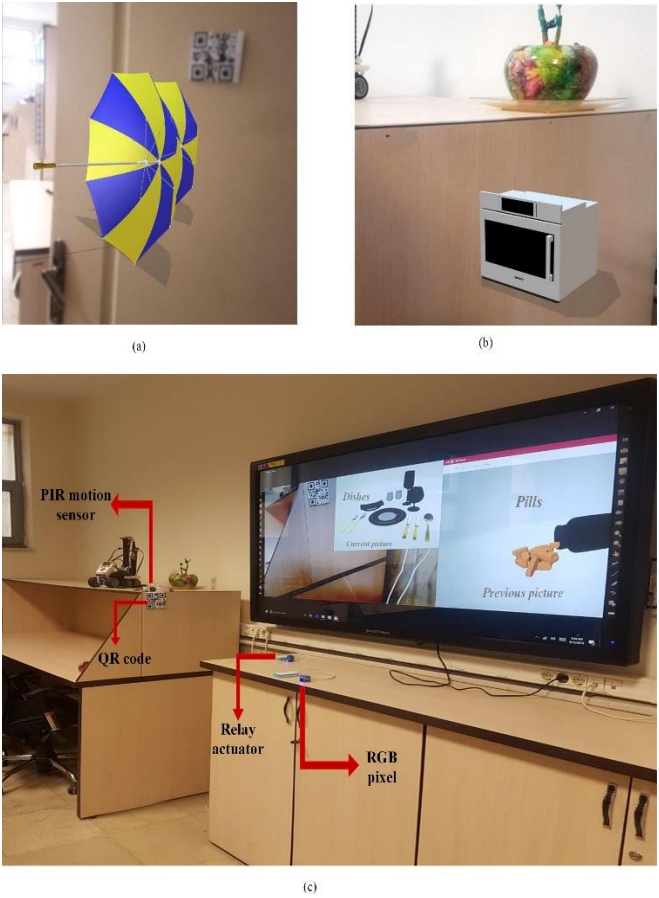

Fig. 3. (a) The rain warning was detected by the rain sensor and a picture of an umbrella is generated as a reminder, (b) the flame sensor in the kitchen near the oven detected the presence of a flame, so relay actuator was activated and user received an image notification, (c) an illustration of the user's field of view shown on the LCD screen.

3) If user enters the kitchen and the PIR detects this event, then they can see dishes picture. Moreover, an audio alert is played to reminding them of the place of the dishes.

4) If the temperature sensor detects that the indoor weather is cold, then a text message is shown to the user on the screen. If this message is confirmed by the user, then the heater can be on.

5) If the flame sensor detects a flame or fire, then relay actuator is activated and oven can be turned off.

## V. RESULTS AND ANALYTICS

In this section, we present our AAL system reliability and performance according to effective parameters and preliminary experimental results. First, we consider minimum QR code size that is large enough to be scanned by camera, then we describe system response time in different conditions.

### A. QR code minimum size

The minimum printed size of the QR code is dependent on some factors:

1) Camera parameters
2) Number of modules
3) The distance between the code and scanner
4) Scanning environment quality

In this study, we have used version one of the QR code which has 21*21 modules. Two minimum QR sizes are defined based on first, the environmental parameters and the second one, the camera parameters. The final QR code size is determined according to these calculated parameters.

The first minimum QR length is calculated as

$$Minimum\ QR\ Code\ Size\ 1: L_{min1} = \left(\frac{D_{scan}}{K_{dis}}\right) * K_{den} \quad (1)$$

Where $K_{den} = \frac{21}{25} = 0.84$ is the data density factor (the maximum module number divided by 25 to normalize it to the equivalent of a Version 2 QR); $K_{dis} = 10$ the distance factor start from a factor of 10 reduced by 1 for each of poor lighting in the scan environment, a mid-light colored QR code being used, or the scan not being done front on; and $D_{scan}$ defined 300mm as maximum scanning distance.

The second minimum QR size is calculated as following:

$$Minimum\ QR\ Code\ Size\ 2: L_{min2} = \frac{PPQ * FOV}{CCD_W} \quad (2)$$

Where FOV is the camera's field measured by experiment, in this case 340 mm; PPQ is defined as pixels in each dimension needed per QR. Considering 10 pixels needed for each module we define:

$$Pixel\ per\ QR = PPQ = 10\ pixels * 21\ modules \quad (3)$$

And $CCD_W$ is the width of the CCD array calculated by solving the following equations:

$$Camera\ Resolution: CCD\ Area = CCD_W * CCD_H \quad (4)$$

$$CCD\ Width\ and\ Height: CCD_W = \phi * CCD_H \quad (5)$$

Where ϕ is normally the golden ration defined as:

$$Golden\ Ratio = \phi = \frac{1 + \sqrt{5}}{2} \cong 1.618 \quad (6)$$

Finally, we can find the minimum required QR size for printing according to the following logic:

$$Minimum\ QR\ Code\ Size: L_{min} = \max(L_{min1}, L_{min2}) \quad (7)$$

Based on our 12MP camera and computational result as shown in equations, we must prepare at least 21*21mm printed QR code.

### B. Response time

For estimating the system performance and complexity, we determined application computational response time and analyzed battery usage. For example, we first published a new value for turning on relay actuator via MQTT protocol, then time response for playing an audio message was assessed. In another test, we scanned a QR code and response time for displaying an augmented image, after publishing a new message for turning off the relay actuator via MQTT, was measured. Both experiments were then replicated under the same conditions fifty times to evaluate system accuracy as shown in Fig. 4 and Fig. 5. The system required an average 364ms to play an audio message and for displaying three-dimensional image based on QR code detection, the average response time was 106ms. We found that playing audio message instead of displaying image message, had better performance and less battery usage. However, using QR code to display AR image could decrease the response time in compare to playing an audio message.

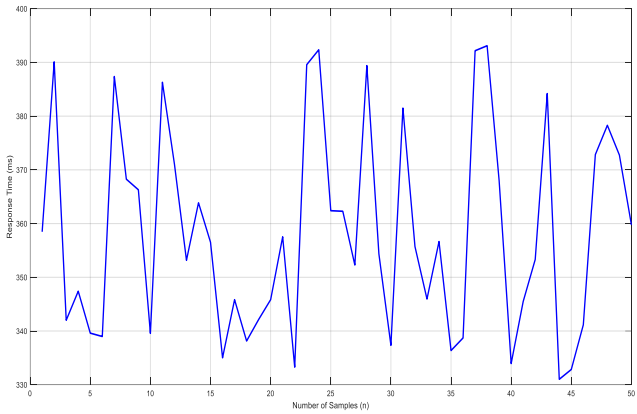

Fig. 4. Voice message response time after publishing a value to the MQTT server.

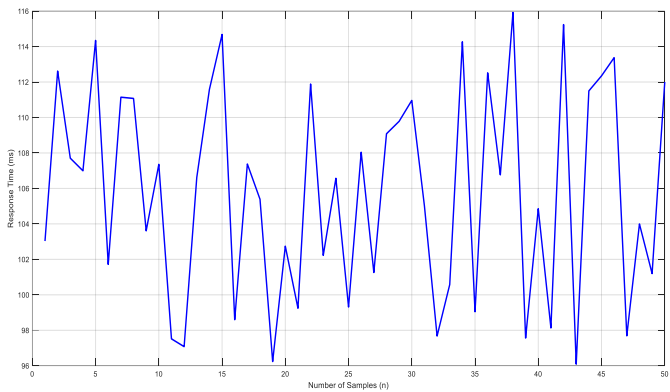

Fig. 5. Image message response time after publishing a value to the MQTT server.

We were aware that the average publish-subscribe latency in MQTT server is estimated 120ms and data-transfer rate is limited, so our algorithm executed in an accurate way according to the results. For real-time systems, similar to our work, this delay was adequate. Moreover, data loss did not occur after performing a series of data transfers.

## VI. CONCLUSION AND FUTURE WORKS

We have presented the AAL system, which uses ARCore library for Android operating system to assist people suffering from AD and make them live more independently through interaction. This system can be used by mild to moderate AD patients those who perceive their disease as a memory impairment. We have also designed Android and windows application, which allow caregivers and family to monitor patient via IoT platform, so they can be notified with the all the environment sensors values. In addition, they are able to control actuators states, for example to remind patient medication time.

We conducted an evaluation of AAL system to analyze its performance under the several conditions. We found that playing audio message instead of displaying image message, had better performance and less battery usage. However, using QR code to display AR image could decrease the response time in compare to playing an audio message. The result showed that MQTT is adopted for a quick and reliable messaging transport among various devices.

In this paper, we performed operational test and evaluation of the system. Our current plan is to implement our system on Google Glass for real scenarios of AD patient daily life who already uses paper tags placed in home by family.

ACKNOWLEDGMENT

This work was supported by the Cognitive Sciences and Technologies Council of Iran.